# Transformer Based Machine Fault Detection From Audio Input

Kiran Voderhobli Holla

*Abstract*—In recent years, Sound AI is being increasingly used to predict machine failures. By attaching a microphone to the machine of interest, one can get real time data on machine behavior from the field. Traditionally, Convolutional Neural Net (CNN) architectures have been used to analyze spectrogram images generated from the sounds captured and predict if the machine is functioning as expected. CNN architectures seem to work well empirically even though they have biases like locality and parameter-sharing which may not be completely relevant for spectrogram analysis. With the successful application of transformer-based models in the field of image processing starting with Vision Transformer (ViT) in 2020, there has been significant interest in leveraging these in the field of Sound AI. Since transformer-based architectures have significantly lower inductive biases, they are expected to perform better than CNNs at spectrogram analysis given enough data. This paper demonstrates the effectiveness of transformer-driven architectures in analyzing Sound data and compares the embeddings they generate with CNNs on the specific task of machine fault detection.

## I. INTRODUCTION

Machine failures lead to unplanned downtime, loss of productivity, and hurt brand value. Predictive Maintenance therefore assumes utmost importance in Industry 4.0. With the proliferation of IOT, various types of sensors can now be deployed along with the machine of interest and real time feedback can be obtained from the field which enables fault predictions in advance. An interesting addition to the long list of potential sensors is the rather simple and ubiquitous microphone. At an extremely low cost, it can collect various sounds emanating from the machine of interest and transmit it back to a central server for analysis and fault prediction.

Sound based fault prediction systems have significant advantages over alternate mechanisms. Apart from the low cost of the sensor and the easy installation, these are essentially contact-less i.e., the sensor only needs to lie in the vicinity of the machine and need not be physically attached to the same. Thus, it is completely non-intrusive and there is no risk of the sensor interfering with the (usually sensitive) equipment's performance or posing a danger to it. An even bigger advantage is that these usually provide lead indicators of faults unlike other sensors which usually provide lag indicators. For e.g., the microphone could pick up anomalous vibration sounds from the machine well before it begins to heat up and starts activating the temperature sensors. In this context, the terms 'fault prediction' and 'fault detection' can be used interchangeably. Lastly, the ability to detect anomalies in machine sounds is a well-accepted art that comes with years of practice and domain experience [1]. It is now possible to train Machine Learning models at replicating this art.

Training Machine learning models for fault prediction leveraging Sound AI has been well explored in recent years. Usually, Fast Fourier Transforms (FFT) are used to decompose the audio signal into its frequency components by applying the Fourier transform to successive overlapping signal segments. The result is a spectrogram, which is a two-dimensional representation of the signal in the time–frequency domain. With the proliferation of pre-trained models and the success of transfer learning, it became possible to leverage pre-trained CNNs by feeding them spectrograms images and fine-tuning them for a downstream task like fault detection. Alternatively, unsupervised learning techniques could be used, and this involves extracting features from the first model and feeding them to another downstream model which leverages unsupervised anomaly detection techniques to isolate faulty sounds (Fig 1).

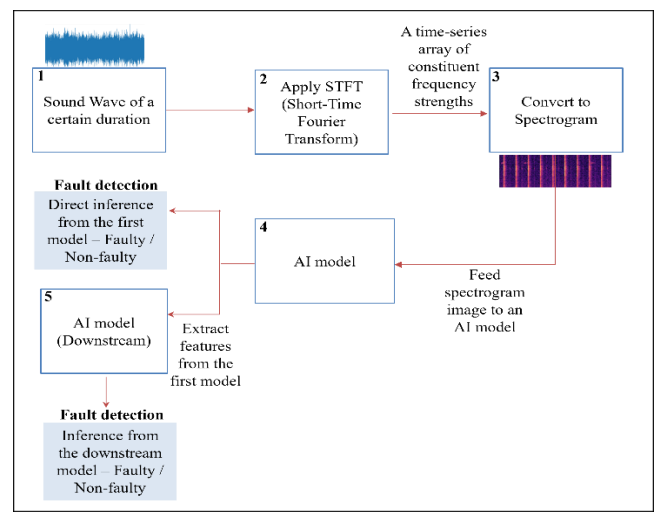

Fig. 1.  Popular techniques for fault detection using Sound AI

While CNN models are efficient and seem to work well empirically with spectrogram images, they have some characteristics which may hinder an optimal performance. CNN architectures, by their very nature, enforce inductive biases. The weights of CNN models are constrained by locality and weight-sharing. However, spectrogram axes represent frequency and time unlike a conventional image. An object at any location in a conventional image is still the same object. However, a pattern at the bottom of a spectrogram is different from a similar pattern at the top due to frequency being on the y-axis. Moreover, there could be long range dependencies which are not well captured by a CNN. More recently, transformer [2] based architectures have been successfully used in Vision problems starting with Google's Vision Transformer (ViT) 2020. Since transformer-based architectures have significantly lower inductive biases, they are expected to perform better than CNNs at spectrogram analysis given 'enough data'. They may also benefit from capturing long range dependencies.

The challenge then is to gather 'enough data' in order to train and make the transformer model effective for Sound classification tasks. The lack of such training data prompted the Audio Spectrogram Transformer (AST) [3] team to leverage ViT weights as a starting point, since ViT was already trained on millions of images. While, AST had a very similar architecture as ViT, care had to be taken to transfer 2D spatial information of positional embeddings from ViT to AST so that AST could use the pretrained ViT weights for initialization and then further train specifically for sound classification. The AST team achieved state-of-the-art results on AudioSet, ESC-50, and Speech Commands classification tasks.

This paper aims to evaluate the effectiveness of a transformer-based architecture over a conventional CNN-based architecture on the sound data domain. We leverage a public dataset titled Malfunctioning Industrial Machine Investigation and Inspection (MIMII) [4] having labelled data of both normal and faulty machine sounds. MIMI was the first public dataset containing sounds of industrial machines under normal and anomalous operating conditions in real factory environments. It was released with the aim of assisting the machine-learning and signal processing community to enhance the field of automated facility maintenance. It continues to remain a popular choice for benchmarking in the research community particularly in the domain of machine sounds. The AST architecture is used as the choice of the transformer model as it is amongst the first convolution-free, attention-based model for audio classification. The utility of the embeddings generated by the AST-based model is compared with those generated by a CNN model of equivalent size. ResNeXt (Residual Networks with Aggregated Transformations) [5] is used as the choice for the CNN model. This model out-performs the older ResNet models and can be considered as the state of the art. In particular, resnext101_32x8d from the PyTorch Image Models (timm) repository is selected as this has 87 Mil parameters and is comparable in size to the AST model.

The pretrained AST model is first fine-tuned using supervised learning techniques to generate (transformer) embeddings. We then fine-tune the CNN model on the same dataset to generate CNN embeddings. In both cases, the fine-tuning is done by leveraging normal as well as anomalous samples from the MIMII dataset. The fine-tuned AST and CNN models are then compared at the task of fault detection. In practice, anomalous machine sounds are rare to capture and therefore access to labelled anomalous data may not always be guaranteed. Hence, embeddings generated from the non-finetuned models of CNN and AST are also captured and used on a downstream non-supervised task of sound anomaly detection. The results are then compared to see if the transformer model performs better due to the lower inductive bias inherent in their architecture. This paper also analyzes the attention patterns in the internal transformer layers and the gradient-weighted Class Activation Mappings in the CNN to understand the reasons behind the performance.

## II. RELATED WORK

Sound AI techniques have been proposed to diagnose faults in diesel engines [6], induction motors [7], gearboxes [8] and other similar equipment. The first step usually involves extracting features that capture the temporal and spectral properties of the input sound. These features are then fed to a subsequent decision module which detects out-of-distribution samples. In classical machine learning, techniques such as a 1-class classifier support vector machines or distance-based methods [9] or density-based methods such as the Local Outlier Factor (LOF) [10] are used to detect out of distribution samples which are regarded as faulty.

More recently, a class of deep neural networks called autoencoders have been employed to detect machine anomalies. These are essentially reconstruction-based methods and the fundamental idea behind these methods is that a normal sound signal can be reconstructed accurately from a reduced latent space representation, whereas anomalous sounds cannot be reconstructed without a significant degradation. For e.g. [11] employed a neural network with an autoencoder structure to detect abnormalities in the emitted sound of a surface-mount device. Autoencoders are trained to reconstruct regular samples by learning an efficient representation of the input vector. The reconstruction residual-error is used as a metric to detect machine faults. These kinds of models can be trained using only the normal machine sound data and are particularly suitable for anomaly detection, where the volume of anomalous sound data is much smaller than normal sound data.

In classification-based approaches using deep learning, supervised anomaly detection models are used which necessitate the presence of sufficient anomalous sound data during the training phase. In this approach, a classifier is trained from a set of labelled data instances containing normal as well as anomalous sounds, and the learned model is then used to classify test instances.

Since anomalous sound data is almost always in short supply, an interesting two-part hybrid approach gaining popularity in past couple of years is to feed the sound data to a pre-trained deep learning model like the CNN or a Transformer and use the embeddings generated by the model's penultimate layer as input features for another downstream model which uses traditional techniques like distance or density-based information to identify anomalies. Additionally, fine-tuning on auxiliary tasks in the same domain has been shown to strengthen the embeddings generated from the first model. For e.g. in [12], the dataset used had sound samples from variety of industrial machines and the authors finetune a Vision Transformer (ViT) on the auxiliary task of classifying the machine-type based on the sound sample. Though the fine-tuning task was unrelated to anomaly detection, it enhances the discriminative power of the embeddings generated by the ViT model. These embeddings were then fed to another downstream model employing k-Nearest-Neighbour distance-based techniques to predict if the sound sample is anomalous or not. Thus, the first model essentially acts as a "feature extractor" and the embeddings it generates contain useful information learnt during pre-training and subsequent fine-tuning which a downstream model can make use of.

While there have been studies leveraging transformers pre-trained on image inputs like ViT, the usage of transformers trained on audio inputs like the AST is a relatively new and emerging area. This study goes a step further and compares the effectiveness of embeddings generated from an AST transformer model as compared with the embeddings generated from a similar sized CNN model on a downstream task of sound anomaly detection. The main focus in this work is the comparison of the effectiveness of CNN and AST embeddings generated from various types of sound inputs to detect anomalous samples, which to the best of my knowledge is the first such attempt. This paper also dwells deep into the inner layers of each model to understand the reasons behind the performance of the embeddings generated. In DCASE 2023[13], the two-part hybrid approach outlined above was extensively used in many winning solutions and this approach can indeed be said to be the state of the art as of today. The success of the two-part hybrid approach relies excessively on the discriminative power of the embeddings that are extracted from the first model and hence it is vital to explore, compare and analyze the embeddings extracted from different classes of models as done in this study.

### III. Dataset

The MIMII dataset has four machine sound types—(i) valves, (ii) pumps, (iii) fans, and (iv) slide rails—and for each type of machine, they provide sounds from multiple product lines.

All sounds are of 10-second segments and comprise of normal as well as anomalous sounds. The sounds are captured from 8 microphones kept at a distance of 50 cm from the machine in a circular fashion. The sound is sampled 16 kHz. Apart from the target machine sound, background noise in multiple real factories was continuously recorded and later mixed with the target machine sound. Sound samples with signal to noise ratio (SNR) of 6 dB are used in this study.

The valves are solenoid valves that are repeatedly opened and closed. The pumps are water pumps that drain water from a pool and discharge water to the pool continuously. The fans represent industrial fans, which are used to provide a continuous flow of gas or air in factories. The slide rails in this paper represent linear slide systems, which consist of a moving platform and a stage base.

The MIMII dataset ensures that different real-life anomalous scenarios have been considered for each kind of machine: for e.g., contamination, leakage, rotating unbalance, rail damage, etc (Table I). Additionally, background noise from real factories was mixed with the target machine, thus ensuring that the resulting sounds are as close to real life factory environments as possible.

TABLE I
LIST OF OPERATIONS AND ANOMALOUS CONDITIONS

| Machine Type | Operations | Examples of Anomalous conditions |
|---|---|---|
| Fans | Normal operation | Unbalanced, voltage change, clogging, etc. |
| Valves | Open / close repeat with different timing | More than two kinds of contamination |
| Pumps | Suction from / discharge to a water pool | Leakage, contamination, clogging, etc. |
| Slide rail | Slide repeat at different speeds | Rail damage, loose belt, no grease, etc. |

However, there are couple of limitations which need to be considered:

- Domain shifts caused by changes in state of the machine or changes in the environment are possible and hence the models trained on such datasets need to have domain adaptation capabilities to perform well in inference environments. Irrespective of the choice of the model, domain adaptation techniques [14] can be employed to ensure that the model adapts to the target environments seamlessly.

- A second limitation of the dataset is that it is imbalanced and contains considerably less abnormal data than normal data. To mitigate this, we use the Area Under the Curve (AUC) as the evaluation metric since this is sensitive to class imbalance. Moreover, in addition to supervised approach, we also evaluate the effectiveness of models using an unsupervised approach which does not need anomalous samples during training.

The baseline code provided with the MIMII dataset consists of a simple autoencoder that uses normal sounds for training and anomalous sounds (along with unseen normal samples) for the testing phase.

TABLE II
SOUND SAMPLES USED IN THIS WORK

| Sound Sample | Normal Sound filecount | Anomalous filecount |
|---|---|---|
| fan_00 | 1011 | 407 |
| fan_all | 4075 | 1475 |
| valve_all | 3691 | 479 |

| pump_all | 3749 | 456 |
| slide_all | 3204 | 890 |

25% samples are taken for the test phase which remain untouched till inference. In the remaining samples, 10% are kept for validation and the rest for training. In unsupervised approaches, the anomalous samples from train and validation datasets are simply discarded.

Among the machine types, only 'fan' has a fair number of anomalous samples. A very basic version of all models is tuned on the training/validation data of this machine (fan with id_00, henceforth called fan_00 set). The hyper-parameters selected are then used to train and infer each model separately using all product lines for that machine at a time. Thus, when dealing with valves, training and inference is done for all product lines under valve (which combines samples from id_00, id_02, id_04, id_06).

## IV. PROPOSED APPROACH

Three models are used in the study – the autoencoder baseline which is provided as part of the MIMII dataset, the CNN model and the transformer model. The audio pre-processing steps are outlined first followed by each model's architecture description, loss function and the evaluation function.

For the pre-processing requirements, I mainly use the Librosa [15] package. The audio data in the MIMII dataset is in the form of WAVeform audio file (.wav) format. The audio is spread across 8 channels recorded from eight distinct microphones. This audio data needs to be converted to a single channel so that it can be processed further. The signals could be averaged across channels for each sample or alternatively, the sound from one of the microphones could be taken forward for processing while ignoring the other channels. For the objectives of this study, both approaches lead to similar results. I used the averaging option. The MIMII dataset was recorded at a sampling rate of 16,000 (Hz) and each single .wav file is a segment of 10 seconds. In order to ensure that this signal can be processed by the models chosen, it needs to be converted to an appropriate input format using a set of pre-processing steps. Log-Mel spectrogram is the choice of input format for the models. Compared to a classic spectrogram, a Mel-spectrogram simulates the sound perception of humans and are a popular choice.

Log Mel-spectrograms are generated from the input signals as follows:
1) The audio signals are mapped from the time domain to the frequency domain by repeatedly applying Fourier transforms. This is done in an overlapping manner across the entire length of the signal. The output is a spectrogram.
2) The spectrogram is further updated by applying the Mel scale to the frequency (y-axis) and the color dimension (amplitude) to decibels. The Mel scale is a quasi-logarithmic function of acoustic frequency designed such that perceptually similar pitch intervals appear equal in width over the full hearing range.

The key parameters involved in this process are the sampling rate, number of Mel bins (n_mels), number of samples per frame (n_fft) and the frame stride (hop_length) which determines the extent of overlap. Librosa is used for generating log Mel-spectrograms for the Autoencoder and the CNN model. A sampling rate of 16,000 is used which is the rate at which input audio samples in MIMII are recorded. Default values for other parameters in Librosa are used unless otherwise noted. There are minor changes in pre-processing based on the individual model as outlined below.

- For the default Autoencoder, log Mel spectrograms are generated using n_fft length of 1024, hop_length of 512 and 64 n_mels. This generates spectrograms of size 64 Mel energies over 313 time slices. In line with the MIMII baseline, a sliding window context of size 5 frames is used to capture change in energies. This results in 309 samples for every sound sample, each having 64*5 Mel energies which gives the final shape of <309, 320>. Each of these 309 samples is considered as a separate training sample. Inference is done by simply making the batch size as 309, taking the average reconstruction error over a batch to be the output for that single sound sample and then measuring the deviation from the original signal. It must be noted that I have deliberately not deviated from the original MIMII baseline.

- For CNN model, hop_length of 128 is used and the number of Mels (n_mels) is set to 128. The generated output is normalized, then triplicated across 3 channels and the resulting <3, 128, 1251> shaped output is fed to the CNN model. The triplication is needed since the CNN model of choice accepts 3-channel data as input.

- In case of the AST model, the AST library-provided 'feature extractor' is used to directly generate the log-Mel spectrograms from the raw sound file. This library uses a sampling rate of 16,000 and n_mels = 128 by default, generating log-Mel spectrograms of shape <128, 1000> for a 10 sec audio file which is comparable to the dimensions of the spectrogram generated for the CNN model above.

The details of the three AI models used in the experiments are explained below:

*A. Autoencoder (AE) Model description*

The MIMII dataset is accompanied by a simple autoencoder baseline which is used for unsupervised fault detection. The spectrogram samples are passed through an encoding section of Fully Connected (FC) layers. The data then feeds into a bottleneck layer with a small number of nodes which forces the network to compress the input signal into the lower dimensional representation. This compressed representation feeds into a decoding section mirroring the same architecture as the encoder section in order to recreate the input signal. The dimensions of each FC layer are as follows:

- *FC layer 1 (FC1): Input dimensions (i/p dim) = Number of dimensions (num_dim) in original(orig) signal. Output dimensions (o/p dim) = 64*
- *FC2: i/p dim = 64. o/p dim = 64*
- *FC3: i/p dim = 64. o/p dim = 8*
- *FC4: i/p dim = 8. o/p dim = 64*
- *FC5: i/p dim = 64. o/p dim = 64*
- *FC6: i/p dim = 64. o/p dim = 64, num_dim in orig signal*

This model is trained using only normal signals. The loss used is a Mean Squared Error between the original signal and the signal output by the model. Thus, the model is motivated to compress and decompress the signal retaining as much information from the original sample as possible. When an anomalous signal is fed to the model, the reconstruction error will be much larger than expected from the training set because the signal compression and decompression scheme learned by the network is only expected to work well for normal samples.

The autoencoder model provided in MIMII baseline is rewritten in Pytorch and used without further fine-tuning as improving the baseline is not the primary focus of this paper.

*B. CNN Model description*

As outlined earlier, resnext101_32x8d model from the PyTorch Image Models (timm) repository is used. The input to this model is the preprocessed spectrogram outlined in the previous section. This model has multiple convolution blocks one after the other and the final block outputs an embedding of 2048 dimensions.

In the unsupervised approach, these embeddings are extracted as-is and fed to a downstream model which does the anomaly detection. Scikit's Local Outlier Factor (LOF) algorithm is chosen as the downstream model. It computes the local density deviation of a given data point with respect to its neighbours and considers as outliers those samples that have a substantially lower density than their neighbours'. The fitting and inference of this model are near instantaneous even on a CPU and hence this was the preferred choice.

In the supervised approach, a sequential module with the below layers is added as the head of the model to enable classification.
- *Linear(in_features=2048, out_features=256, bias=True)*
- *Dropout(p=0.1, inplace=False)*
- *ReLU()*
- *Linear(in_features=256, out_features=1, bias=True)*

This model is finetuned using both normal and anomalous samples and then used for inference. In order to regularize the model, the number of parameters are reduced by freezing the lower layers of the CNN and making only the upper few layers trainable. The Learning Rate (LR) is progressively reduced across each 'trainable' layer of the CNN architecture moving backwards from the head. Thus, the fully connected layers are trained at the rate of LR, the block prior to that (bottleneck layer 3 of the CNN) is trained at LR/10, the block prior to that (bottleneck layer 2) trained at LR/100 and so on. Thus, the head, which has untrained weights makes maximum movements based on the gradients, while the prior layers make smaller movements. At lower layers, weights are completely frozen since it is expected that the model can make full use of the pretrained weights without needing to recalibrate them for the classification task at hand. Empirically, these techniques seem to help in regularizing the model as well as making it parameter efficient. This makes it easy to experiment with large models on a frugal infrastructure.

*C. AST Model description*

Similar to CNN, the following approach is taken.
- In the unsupervised approach, the embeddings are generated by simply taking the output values from the penultimate layer of the pretrained, non-finetuned AST model and then feed it to the downstream LOF model for anomaly detection.
- In supervised approach, the same head structure (outlined earlier) is appended to the AST model. The head has untrained weights like in the CNN. Here too, parameter freezing is done for the lowest layers of the AST. The head and a few upper layers of the AST are trained with different learning rates as explained in the CNN section. The LR is thus decayed layer upon layer moving backwards from the head and this enables fine-tuning on a frugal infrastructure.

The loss functions vary based on the model and are outlined below along with other training details:
- Mean Squared error for the reconstruction errors in the Autoencoder. The Epoch count is 50 and the LR is set to 0.001 with Adam optimizer.
- Binary Cross Entropy (BCE) is used as the loss function for the supervised approaches in both CNN and AST. One training epoch with an LR of .001 using Adam optimizer is sufficient to bring in a good convergence.
- Default metric for distance measurement (Minkowski) is used in the Scikit LOF library. This requires the setting of the 'contamination' and the 'n_neighbors' parameters. The latter is set to 4 samples as this seemed a reasonable sized neighborhood, however the 'contamination' parameter needs hyper-tuning which is not possible in the absence of anomalous data. This is because in the unsupervised approach, all anomalous data samples from train and val are discarded. To solve this conundrum, I simply use contamination values from 0.1 to 0.4 during training and just take a simple majority vote over the predictions during inference.

The evaluation function is common for all models and is the AUC (Area Under Curve).

V. EXPERIMENT RESULTS

Table III contains the AUC across models.

TABLE III
AREA UNDER CURVE

| Model | Fan: 00 | Fan - All | Pump - All | Valve - All | slide - All |
|---|---|---|---|---|---|
| Autoencoder Baseline | 0.7770 | 0.9081 | 0.8720 | 0.5757 | 0.8684 |
| CNN - Supervised | 0.9990 | 0.9721 | 0.9874 | 0.9884 | 0.9990 |
| AST - Supervised | **0.9990** | **0.9779** | **0.9904** | **0.9999** | **1.0000** |
| CNN - LOF | 0.7478 | **0.8055** | 0.8201 | 0.7761 | 0.8144 |
| AST - LOF | **0.7716** | 0.7696 | **0.8502** | **0.8099** | **0.8367** |

## VI. OBSERVATIONS AND ANALYSIS

The proposed approaches are objectively evaluated using the Area Under the receiver operating characteristics Curve (AUC) metric for each model. The performance of the CNN-based and the AST transformer-based models are compared against each other in both the supervised and the unsupervised anomaly detection approaches. The results for each machine type are shown in Table III. Row 1 contains the results from the baseline model. Rows 2,3 show the AUC of the supervised approach for the CNN and AST based models respectively. The last 2 rows show the AUC for the unsupervised approach.

Before analyzing the results, it is useful to analyze the pre-processed spectrogram images which are fed as input to the AI models. Fig 3 shows normal sound spectrogram and Mel-spectrogram for the fan (top) and the valve (bottom). There are differences due to the nature of operations of the machines. The fan and the pump are stationary, and their spectrograms differ from the valves and the sliders which are non-stationary machines. The patterns in the latter are more pronounced and impulse-like whereas the former has more continuous patterns with lesser spikes which might be more difficult for the AI model to analyze.

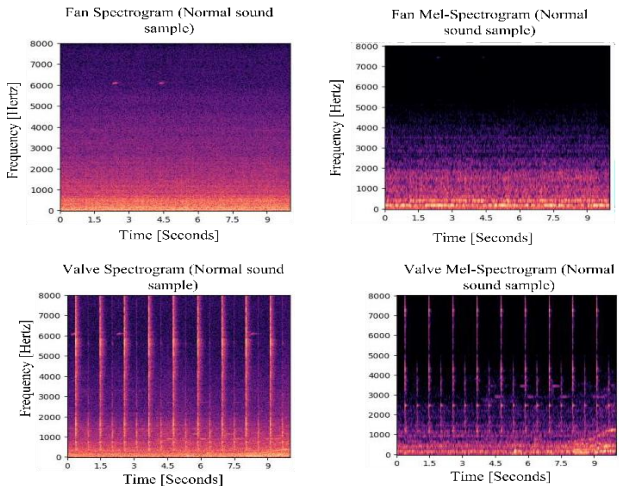

Fig. 2. Spectrogram and Mel spectrogram for a normal sound sample of fan (top) and Valve (bottom)

In general, the results show that the AST transformer-based model outperforms the CNN-based model in the supervised learning scenario. Even with very few anomalous samples, both models converge in just one training epoch and are able to classify sounds with a fair degree of accuracy. Among the different types of machines, it can be seen that the slider has the highest AUC while the fan's AUC is relatively lower. This is in agreement with other studies which have analyzed the MIMII dataset. As outlined earlier, the spectrograms for sliders have clear visible patterns and provide a rich feature set which can be discerned by the model and aid in its decision making. More importantly, as seen in Fig 3, the difference in normal and anomalous spectrograms are more pronounced for sliders which further helps the model.

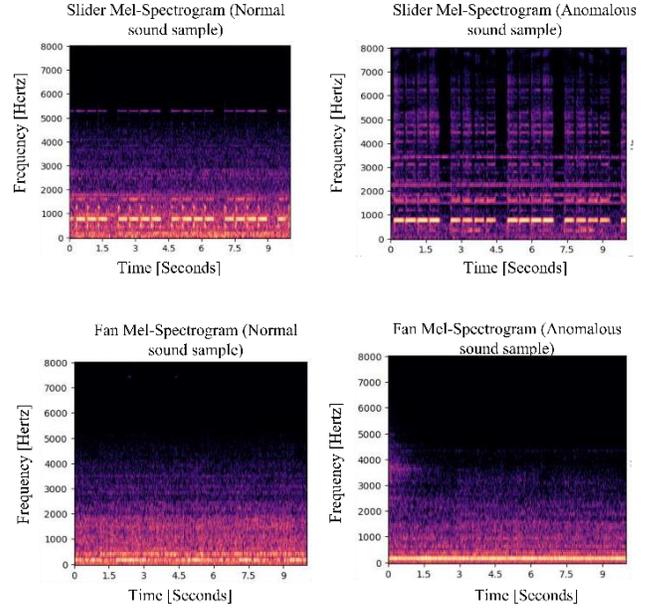

Fig. 3. Mel-Spectrograms for normal(left) and anomalous(right) sounds for slider (top) and Fan (bottom)

While supervised training experiments are not the primary focus area in this study, the intent here is to show that even with very few anomalous training samples, a pretrained transformer such as AST could be fine-tuned in just 1 epoch to give very good fault predictions bettering that of CNN-based models. With the proliferation of digital twins, it could get easier to capture anomalous data over time and hence the supervised approaches to fault detection could start gaining importance.

In the unsupervised task, it can be observed that the transformer architecture again outperforms CNN on all machine types except 'fans'. On manually inspecting the Mel-spectrogram images, it can be reasoned that the likely cause could be the lack of difference between spectrogram images of anomalous and normal fans as compared to other machines which have more fine granular variations and sharper differences. The CNN is able to discern and extract better feature embeddings in such a situation.

To summarize, it is observed that the transformer model performs better than the CNN in the supervised approach and the embeddings it generates are slightly better than the CNN embeddings in the unsupervised approach as well. By using t-SNE plots [16], we can visualize the embeddings generated in a two-dimensional space as done in Figure 3.

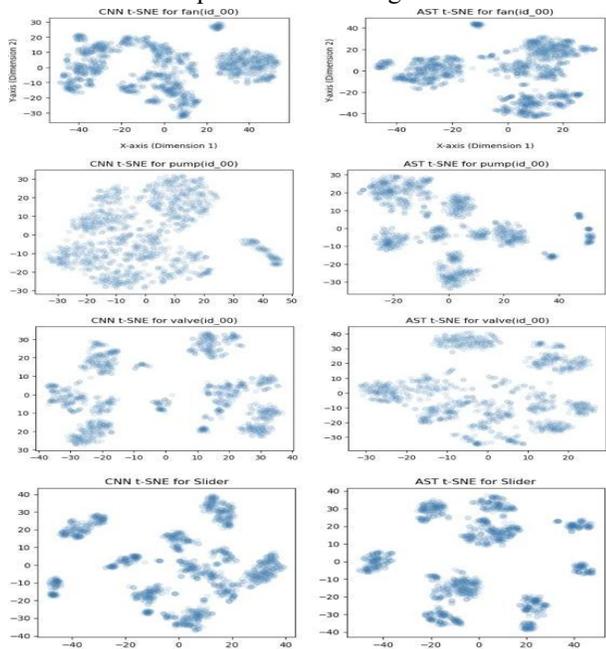

**Fig. 3. Visualization of CNN/AST embeddings in a two-dimensional space using t-SNE**

It is immediately noticeable that the AST transformer embeddings cluster the sounds into well-defined spaces relatively better than the CNN for most machine types. It is thus able to isolate and classify failures relatively better than the CNN as shown in Figure 4. These results are further enhanced after fine-tuning but what is important to note is that even without fine-tuning, the performance of the transformer-based model is relatively better.

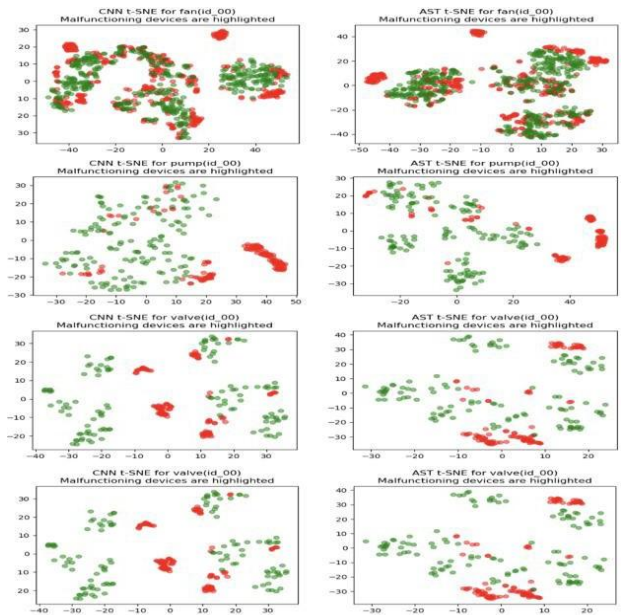

**Fig. 4. Embedding visualization highlighting faulty sounds in a two-dimensional space using t-SNE**

The main advantage of a transformer model is its ability to effectively capture global contextual information through the self-attention mechanism, enabling it to model long-range dependencies and contextual relationships, which can improve robustness in tasks that require understanding global context. Also, the weights of CNN models are constrained by locality and weight-sharing. These inductive biases are hard-wired into the CNN model due to its architectural structure and indeed may be responsible for its success as far as traditional images are concerned. By encouraging translation equivarance (without pooling layers) and translation invariance (with pooling layers), the convolutional inductive biases make CNN models more sample efficient and parameter-efficient [17]. However, both these biases could hamper its performance as far as spectrogram images are concerned:

- **Weight sharing bias:** The use of pooling layers common in CNNs along with the inductive bias of weight sharing is used to achieve translational invariance. This means a particular pattern of pixels is recognized wherever it is in the image and treated in a similar fashion by the model. This works well for regular images but in case of spectrograms, the axes represent frequency and time and thus a pattern at the bottom of a spectrogram is different from a similar pattern at the top due to frequency being on the y-axis. Such patterns shouldn't be treated the same way by the model and weight sharing across different locations could potentially hamper the model from discerning discriminating features which could help identify anomalies. Transformer based architectures do not suffer from this inductive bias and hence tend to provide a better feature set as far as audio inputs are concerned.

- **Locality bias:** The locality bias of CNNs impair its ability to capture long-range dependencies. In contrast, transformer architectures like the AST, make use of the self-attention mechanism which is global in nature since it gathers information from the entire image. They are thus able to effectively capture and corelate distant pieces of information in an image. Moreover, unlike the CNNs which build the receptive area gradually across the layers, the representation in transformer architectures is similar in every layer. It thus literally has a "big picture" even in its early layers. This has been proven beneficial even in case of certain types of regular image analysis such as histopathological analysis [18], where it is critical to consider not just the region of interest, but also the neighboring tissues when diagnosing a particular disease. Indeed, Bello et al. [19] study this bias for regular images and mitigate it by augmenting convolutional layers in CNN with attention. In case of spectrogram images, the entire y-axis reflects the frequency bands at play at any instant of time and there could be important dependencies all across the length of the image. It is likely that the CNN model is unable to capture this information.

A look at the attention pattern in Figure 5 confirms this behavior. The x-axis plots the 12 layers of the AST transformer each having 12 attention heads. The y-axis shows the average attention distance in pixels for each head.

It can be seen that right from the first layer, there are certain heads that are "global" and have a long-distance oversight. A CNN by its very design has a limited view in the early layers.

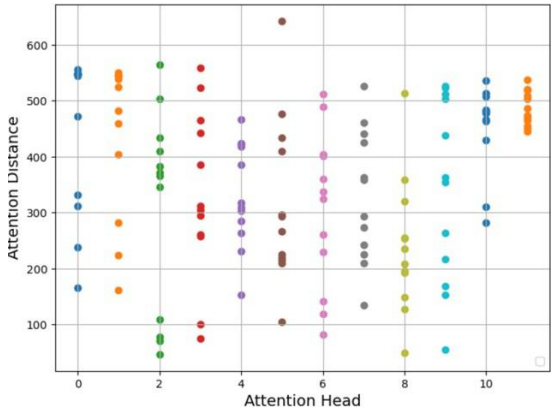

**Fig. 5. Attention patterns in the transformer heads across layers**

In Figure 6, we see the attention pattern of the last layer of the transformer as compared with the Grad-CAM [20] output of the last layer of the CNN. It can be seen that the transformer model is paying attention to the right parts of the input image whereas the CNN model is a bit more diffuse owing to its architecture and the reasons explained earlier. Figure 6 also shows a mid-point representation of the input image across the two models. It is observed that the AST model has well defined attention patterns mid-way itself and knows on what portions of the input spectrogram to focus on. The CNN model on the other hand has still a lot of noise at this stage and needs more layers to be able to focus on the important parts of the input spectrogram. It must be noted that these attention patterns of the AST are, to the best of my knowledge being highlighted for the first time. The original AST paper didn't elaborate on the visualization and explainability aspects.

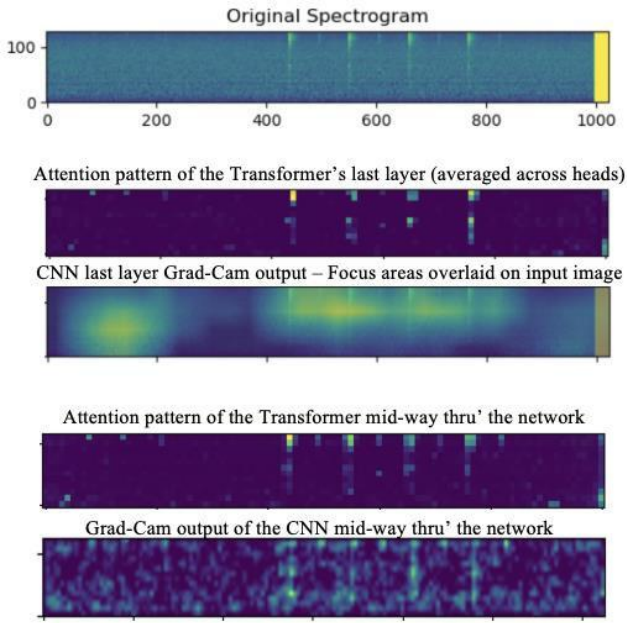

**Fig. 6. Transformer and CNN explainability for an input audio after resizing to the original image shape**

The above results also show that the parameter-efficient training techniques outlined earlier work effectively. Each epoch takes less than 10 mins on a personal M1 MacBook with 16 GB default memory which enables experimenting with fine-tuning large models on sound datasets without additional GPU infrastructure.

## VII. CONCLUSIONS

This study analyzes anomaly detection systems whose inputs are real-life industrial machine sounds. Two anomaly detection systems are evaluated – one based on CNNs, and another based on (AST) transformers. Experimental results show that the transformer-based model clearly outperforms the CNN-based model when using supervised learning methods. The transformer-based model also performs relatively better than CNN-based model when using unsupervised learning methods. Transformer architecture embeddings seem to be better suited for acoustic-related tasks since they have significantly less inductive biases as compared to CNNs, which are constrained by locality and weight-sharing and also lack the ability to capture long range dependencies. To the best of my knowledge, this is the first study which compares and analyzes in detail, the discriminative power of embeddings generated by CNNs and transformers from audio sounds for the downstream task of anomaly detection. This study also explains parameter-efficient techniques to train large transformer models making it easy for future researchers to replicate and enhance this work with minimal infrastructure costs.

The use of Sound AI in field of Predictive Maintenance of machines is the way forward in Industry 4.0. As the number and complexity of the machines increase, it becomes difficult for maintenance workers to use traditional high-cost manual inspection methods. Automatic detection of machine failures through detection models helps cut down on the costs of production downtime [21]. By leveraging the simplest of IOT sensors – a microphone, one can significantly enhance the Prognostics and Health management experience. Sound based anomaly detection systems are contactless, cheap and have a low cost of installation. Since they produce lead indicators of faults, they are of vital importance in mission critical systems where actual failures could be catastrophic.

There are two practical challenges hindering a more wide-spread adoption of these techniques on the field:

- The first challenge is the domain shift usually caused by changes in the state of the machine or changes in the environment itself during inference on the field. The trained models need to have excellent domain adaptation capabilities to perform well in the real-time environment. Domain adaptation techniques such as [14] are a well-researched area and need to be employed to ensure that the model adapts to the target environments seamlessly.
- A second challenge is that the training datasets involved are imbalanced and contain considerably less abnormal data than normal data. Even unsupervised techniques need anomalous data for validation and testing phases, and these are hard to capture. With rapid advancements in generative AI, it may soon be possible to generate realistic, synthetic anomalous data which could be a game changer for this domain. Secondly, digital twin adoption is widely increasing, and this could pave the way for more anomalous data collection from the field which could mitigate this challenge.

## VIII. FUTURE WORK

The era of Large models has begun with parameter sizes running to several billion. Large parameter sizes are associated with the ability to extract and capture more information from the training data thereby creating better feature embeddings. While individual large models could generate embeddings with differing discriminative power in a downstream task, a potential line of future study could revolve around investigating techniques to combine different embeddings in the best possible way in order to enhance the discriminative power of the final feature set. Such ensembling techniques could range from a simple averaging of results across models to more complicated fusion techniques and is a promising area of research. Ensembles with high diversity tend to yield better results [22]. For e.g. the results of this study indicate that embeddings generated from the CNN model are nearly as strong as transformer models in the unsupervised approach. Since transformer and CNN models perform feature extraction in significantly different ways generating diverse feature sets, it naturally follows that an ensemble of the embeddings generated, would likely improve the effectiveness of downstream tasks. It is also possible to enhance the discriminatory power of transformer embeddings in unsupervised settings by simply fine-tuning it on an auxiliary task related to the same domain. For e.g. in [12] the model was first fine-tuned to recognize and classify the machine type based on the input sound data available and then used for feature extraction.

Sound transformers like AST could continue to retain a residual bias as long as transfer-learning from an image-trained model is done. Transformers exclusively trained on sound data from scratch are expected to have far less inductive bias and therefore perform better. Techniques to improve transfer learning from image-based models and/or creating efficient sound transformers from scratch is another future area of study.

While this study has focussed exclusively on machine sound data, the techniques of Anomalous Sound detection used here can be used to enhance the results in other acoustic domains ranging from Conservation & Bioacoustics [23], Safety monitoring [24], Road surveillance systems [25] etc. But perhaps the greatest benefit is in the field of human health and particularly in the field of auscultation. The human body can be considered as one of the most complicated machines on this planet. Predictive health maintenance of the human body is of utmost importance. In particular, leading indicators of health-related faults are needed since failures have catastrophic consequences for the individual concerned. The usage of sound transformers like AST will significantly enhance the results from auscultation performed on patients. Around the time this study was in progress, a paper [26] was released in parallel, where the effect of Abnormal Respiratory Sound Identification using AST Transformer was researched with results surpassing previous state-of-the-art. The usage of transformer-based models in Predictive maintenance using anomalous sound detection techniques will accelerate in the coming days. Efficient embedding extraction and fusion techniques that enhance the discriminatory power of the embeddings will significantly propel this field further.